\documentclass[10pt]{article}
\usepackage[english]{babel}
\usepackage[utf8x]{inputenc}
\usepackage[OT1]{fontenc}
\usepackage{amsfonts, amsmath, amsthm, amssymb}
\usepackage{graphicx}
\usepackage{listings}

\usepackage{multicol} 
\usepackage{amssymb, amsmath, amsbsy}
\usepackage{stackrel}

\usepackage{subcaption}
\usepackage{graphicx}

\usepackage{listings}
\usepackage{csquotes}
\usepackage[hidelinks]{hyperref}
\usepackage[nottoc,notlot,notlof]{tocbibind}

\usepackage[margin=1in]{geometry}
\usepackage{xcolor}
\usepackage{braket}

\title{Relationship between Decimal Hill Coefficient, Intermediate Processes and Mesoscopic Fluctuations }

\author{Manuel Eduardo Hernández-García, Jorge Velázquez-Castro \\
       Facultad de Ciencias Físico Matemáticas \\
       Benemérita Universidad Autónoma de Puebla}

\date{\today}
\begin{document}
\maketitle

\begin{abstract}
  The Hill function is relevant for describing enzyme binding and other processes in gene regulatory networks. Despite its theoretical foundation, it is often empirically used as a useful fitting function. Theoretical predictions suggest that the Hill coefficient should be an integer. However, it is often assigned a decimal value. The deterministic approximation of binding processes leads to the derivation of the Hill function, which can be expanded around the fluctuation magnitude to derive mesoscopic corrections. This study establishes the relationships between intermediate processes and the decimal Hill coefficient through a direct relationship between the dissociation constants, both with and without fluctuations.  This outcome contributes to a deeper understanding of the underlying processes associated with the decimal Hill coefficient while also enabling the prediction of an effective value of the Hill coefficient from the underlying mechanism. This procedure allows us to have a simplified effective description of complex systems.


\end{abstract}

\tableofcontents 

\section{Introduction}
Hill function plays an important role in describing enzyme binding or promoter activity in gene regulation networks  \cite{Alon, Sauro}. However, even if an underlying theory exists for its derivation, it is sometimes used as an empirical description of the biochemical processes. As is the case with the Hill coefficient \cite{Ale}, which is commonly represented as $n$, this coefficient describes cooperativity in the binding of ligands to a receptor molecule, such as enzymes, to a substrate.   In kinetic experiments, data are commonly fitted to a Hill function, and a decimal value is assigned to the Hill coefficient, even though theoretical predictions suggest only integer values are possible.  For example, in a study on the interaction between hemoglobin and oxygen \cite{Hill}, the Hill coefficient was found to be $n=1.8-3.4$. However, theoretical calculations suggest that the value of $n$ should be equal to four because hemoglobin has the capacity to bind up to four oxygen molecules \cite{Motu}. It is evident that this process is more intricate than the ligand-receptor binding described by the Hill function. Consequently, it is employed as a convenient fitting function. In \cite{Kosh, Weiss} intermediate processes were used to provide a more realistic representation of ligand-receptor binding, given that ligands do not always bind simultaneously to a receptor, illustrating how various system parameters can produce different curves with varying degrees of cooperation. However, there is no direct correspondence between the parameters of a Hill function with intermediate processes and those with decimal Hill coefficients.

This study seeks to show that the use of a Hill function with decimal coefficients actually entails the presence of intermediate processes and that the Hill coefficient corresponds to the coefficients of the detailed description that includes the intermediate processes \cite{Weiss}.

 This problem is underexplored in terms of the influence of fluctuations on this type of system. It is widely acknowledged that fluctuations can have consequences in the dynamics of genetic regulatory networks, as demonstrated in references \cite{Scot, Gar}. Although deterministic models can effectively capture most system dynamics \cite{Alon, Scot}, they are insufficient for explaining all the observed behaviors \cite{Thomas}.  In this study, we also investigated the influence of intrinsic fluctuations in ligand-receptor binding on intermediate processes. To achieve this, we employed a systematic expansion around the deterministic behavior to derive mesoscopic corrections \cite{Manuel, Gomez}. Subsequently, we established relationships between the intermediate processes and the decimal Hill coefficient, both with and without fluctuations.  The development of stochastic Hill functions for intermediate processes can improve the quantification of fluctuations in this kind of systems.

 In Section 2, we offer a concise overview of the mesoscopic approximation to study ligand-receptor binding processes that lead to the derivation of the Hill function. Sections 3 and 4 briefly address the deterministic forms of the Hill function with and without intermediate processes, respectively. Section 5 presents the derivation of the Hill function with intermediate processes and stochastic corrections. In Section 6, we establish a connection between the Hill function with decimal coefficient and the Hill function with intermediate processes, with and without fluctuations. Finally, in Section 7, we present our findings and draw conclusions based on the results.

\section{Mesoscopic Approximation}
 It is generally accepted that stochastic models must eventually converge with deterministic models when dealing with large systems. Furthermore, stochastic methods should provide an estimate of the fluctuations in the system. A widely utilized approach is linear noise approximation, which expands the master equation in terms of a small parameter proportional to the reciprocal of the system size \cite{Scot, Gar}. An alternative approach is to expand directly around the mean values of the concentrations \cite{Manuel}, thereby enabling a straightforward method for calculating the mesoscopic corrections to macroscopic dynamics. 

Consider $N$ species $S_j$ ($j$ $\in$ \{$1,2,\ldots,N$\}), and $M$ reactions $\mathcal{R}_i$ ($i$ $\in$ \{$1,2,\ldots,M$\}) such that the species are transformed as
\begin{align}
    \mathcal{R}_i : \sum_{j=1}^{N} \alpha_{ij} S_j \stackbin[k_{i}^{-}]{k_{i}^{+}}{\rightleftarrows} \sum_{j=1}^{N} \beta_{ij} S_j. \label{bd1}
\end{align}
$k_{i}^{+}$ and $k_{i}^{+}$ are the reaction constants. The coefficients $\alpha_{ij}$ and $\beta_{ij}$ are positive integers, from which we find the stoichiometric matrix 
\begin{align}
    \Gamma_{ji} = \beta_{ij}- \alpha_{ij}.
\end{align}
Through collisions (or interactions) of the different elements, they are transformed, so the propensity rates are given as follows \cite{Gar}
\begin{align}
    {t_i^{+}(\mathbf{S})} &= k_{i}^{+} \prod_{j} \frac{S_j!}{\Omega^{\alpha_{ij}}(S_j - \alpha_{ij})!}, &
    {t_i^{-}(\mathbf{S})} &= k_{i}^{-} \prod_{j} \frac{S_j!}{\Omega^{\beta_{ij}}(S_j - \beta_{ij})!}, \label{bd2}
\end{align}
where $\mathbf{S}= (S_1,S_2,..., S_N)$. The propensity rates are important because the master equation is expressed in terms of them. To deduce the evolution of macroscopic quantities from the master chemical equation, the procedure of multiplying it and then averaging is normally employed. This procedure can be used to obtain formal expressions for the equations describing the evolution of mean concentrations and they are written as follows,
\begin{eqnarray}
    \frac{\partial}{\partial{t}} \left(  \frac{\braket{S_j}}{\Omega} \right)= \sum _i \Gamma_{ij} \braket{t_i^{+}(\mathbf{S}) -t_i^{-}(\mathbf{S})}. \label{am}
\end{eqnarray}
 The average reaction rate $\braket{t_i^{\pm}(\mathbf{S})}$ can be approximated using the following expansion around the average \cite{Manuel, Gomez}:
{\footnotesize
\begin{eqnarray}
    \braket{f(\mathbf{X})} \approx &  \left\langle  { f(\braket{\mathbf{X}}) +   \sum_{j_1}  {(\braket{X_{j_1}}-X_{j_1})} \left. \frac{\partial f(\mathbf{X})}{\partial X_{j_1} } \right|_{\mathbf{X}=\braket{\mathbf{X}}} +  \sum_{j_1} \sum_{j_2} \frac{(\braket{X_{j_1}}-X_{j_1})(\braket{X_{j_2}}-X_{j_2})}{2} \left. \frac{\partial^2 f(\mathbf{X})}{\partial X_{j_1} \partial X_{j_2}} \right|_{\mathbf{X}=\braket{\mathbf{X}}}}  \right\rangle \nonumber \\
    =& f(\braket{\mathbf{X}})  +  \sum_{j_1} \sum_{j_2} \frac{\sigma^2_{{j_1},j_2}}{2} \left. \frac{\partial^2 f(\mathbf{X})}{\partial X_{j_1} \partial X_{j_2}} \right|_{\mathbf{X}=\braket{\mathbf{X}}}. \label{2}
\end{eqnarray}}
where $j_1,j_2$ $\in$ \{$1,2,\ldots,N$\}, and $\sigma^2_{j_1,j_2}=\braket{\braket{X_{j_1}}-X_{j_1})(\braket{X_{j_2}}-X_{j_2})}$  is the covariance between variables $X_{j_1}$ and $X_{j_2}$. This approach is a second-order approximation of the mean, assuming that the fluctuations around the mean are small. 
The species mean concentrations are defined as $s_j = \frac{\braket{S_j}}{\Omega}$, where $\Omega$ has units of volume per mole. Therefore, we can rewrite the mean concentration dynamics as
{\small
\begin{eqnarray}
    \frac{\partial s_j }{\partial{t}}=& \sum_{i} \Gamma_{ji}\left( R_i^{D+}(\mathbf{s}) - R_i^{D-}(\mathbf{s}) + \sum_{j_1} \sum_{j_2}  \frac{\sigma^2_{{j_1},{j_2}}}{2}  \frac{\partial^2}{\partial {s_{j_1}}\partial {s_{j_2}}} (R_i^{D+}(\mathbf{s}) - R_i^{D-}(\mathbf{s})) \right), \label{7}
\end{eqnarray}}
where $\sigma^2_{{j_1},{j_2}}= \frac{\braket{\braket{S_{j_1}}-S_{j_1})(\braket{S_{j_2}}-S_{j_2})}}{\Omega^2}$, $\mathbf{s}= (s_1,s_2,..., s_N)$, $R_i^{D+}(\mathbf{s})$ and $R_i^{D-}(\mathbf{s})$ are the deterministic or also called macroscopic reaction rates
\begin{eqnarray}
    R_i^{D+}(\mathbf{s})=& k_{i}^{+} \prod _{j} s_j^{\alpha_{ij}}  , &
    R_i^{D-}(\mathbf{s}) =k_{i}^{-} \prod _{j} s_j^{\beta_{ij}}. \label{8}
\end{eqnarray}
Based on (\ref{2}) and (\ref{7}), we can make the following substitution to obtain a mesoscopic correction to expressions involving deterministic reaction rates, 
 \begin{eqnarray}
     R_i^{D\pm}(\mathbf{s}) \rightarrow  R_i^{D\pm}(\mathbf{s}) +   \sum_{j_1} \sum_{j_2}  \frac{\sigma^2_{{j_1},{j_2}}}{2}  \frac{\partial^2}{\partial {s_{j_1}}\partial {s_{j_2}}} (R_i^{D\pm}(\mathbf{s}) ).  \label{9}
 \end{eqnarray}
Notably, the correction term is proportional to the covariance of the concentrations, which vanishes at the limits of large systems. 

\section{Hill Function with Integer Hill Coefficient}

Hill functions are commonly used to describe enzymatic reactions \cite{Ken} or to capture the dynamics of mRNA synthesis in gene regulatory circuits \cite{Alon} driven by transcription factors. Specifically, the Hill function is employed to describe the stationary concentration of a reversible process \cite{Hill}, as depicted by the following equation
\begin{eqnarray}
    R+nL \stackbin[k_{-}]{k_{+}}{\rightleftarrows} RL_n,
\end{eqnarray}
where $n$ ligands $L$  are bound to the receptor $R$. The concentration of the product $RL_n$ is dependent on the ligands concentrations $l$ and is determined by a Hill function, as expressed by the following Equation (\ref{H1})
\begin{eqnarray}
    H_d(l)= \frac{l^n}{K^n+ l^n}, \label{H1}
\end{eqnarray}
 where $K=\frac{k^{-}}{k^{+}}$ is the dissociation constant and $n$ denotes the Hill coefficient.

\begin{figure} [h!t]
  \centering
\includegraphics[width=.45\textwidth]{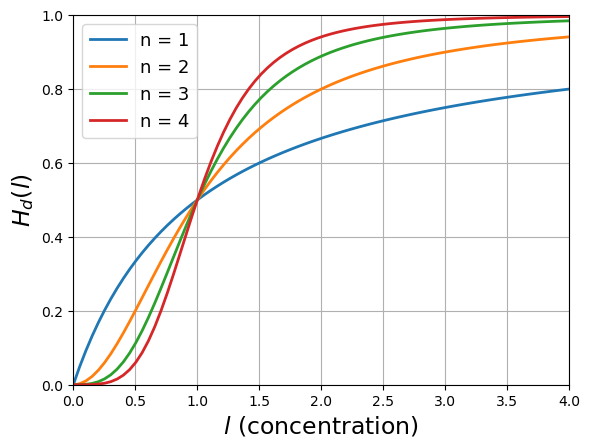}
  \caption{\textbf{Deterministic Hill function.} This figure shows the behavior of the deterministic Hill function for different values of $n$ and $K=1$.}
  \label{figh0}
\end{figure}

Figure \ref{figh0} depicts the Hill function behavior for various values of $n$.  

\section{Ligands-Receptor Reactions with Intermediate Processes}
\begin{figure} [h!t]
\centering
\includegraphics[width=.35 \textwidth]{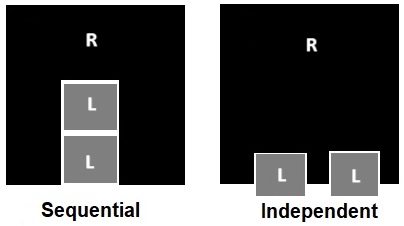}
  \caption{\textbf{Schematization of sequential and independent ligands-receptor binding processes.}}
  \label{fig.1}
\end{figure}

Although Hill derived an equation for integer values of $n$, experimental data demonstrated that the coefficient can assume non-integer values. This is exemplified by the interaction of hemoglobin with oxygen \cite{Hill}, where theoretical predictions estimate a value of $n=4$, yet experimental findings indicate a value of $n=1.8-3.4$. This suggests that Hill's original perspective is insufficient \cite{Ale}. Alternative methods for ligand binding to proteins, such as placing intermediate processes in which sequential or independent binding occurs \cite{Kosh, Weiss}, should be considered. Figure \ref{fig.1} illustrates the potential intermediate processes.

We focused on establishing a steady-state concentration of the product in ligand-receptor reactions involving intermediate processes. These intermediate processes can be either sequential or independent in nature.

\subsection{Sequential case}
In this scenario, the following reactions occur, along with their corresponding reaction constants as reported in \cite{Weiss}
\begin{eqnarray}
    R + n L \stackbin[k_{-1}]{k_{+1}}{\rightleftarrows} RL_{1} + (n-1) L \stackbin[k_{-2}]{k_{+2}}{\rightleftarrows} RL_{2} + (n-2) L ... \stackbin[k_{-n}]{k_{+n}}{\rightleftarrows} RL_{n} , \nonumber
\end{eqnarray}
at the steady state, each chemical reaction must satisfy the equilibrium relationship given by the equation
\begin{eqnarray}
  R_i^{-}=R_i^{+} . \label{5.8}
\end{eqnarray}
This relationship can be equivalently expressed as $1 = \frac{R_i^+}{R_i^-}$, where $R_i^+$ and $R_i^-$ are forward and backward reaction rates, respectively. In this case, the reaction rates are given by
\begin{eqnarray}
    R_i^{+}=& k^{+}_i {l}^{n+1-i} s_{i-1}, \nonumber \\
    R_i^{-}=& k^{-}_i {l}^{n-i} s_{i} ,
\end{eqnarray}
($i=\{1,..n\}$), where we define $s_0$ as the concentration of $R$, ${l}$ as the concentration of $L$ and $s_n$ as the concentration of $ RL_n$.
Using condition (\ref{5.8}), a recurrent relation can be derived from the set of species $s_i$'s, given by
\begin{eqnarray}
    s_{i}=  \frac{ s_{i-1}}{ K_i } l, \qquad i=\{1,..n\}
\end{eqnarray}
($K_i= \frac{k_{-i}}{k_{+i}}$). Solving for $s_{i}$ in terms of $s_{0}$ a relationship between them can be expressed as 
\begin{eqnarray}
    s_i= s_0 {l}^{i} \left( \prod_{j=1}^{i} K_j \right)^{-1}.
\end{eqnarray}
The Hill function, which represents the fraction of the product with respect to all species in the reaction, is then expressed as
\begin{eqnarray}\label{5.9}
    H_d^{S}(l)=  \frac{s_n}{s_0 + \sum_{i=1}^{n} s_i} = \frac{{l}^{n} \left( \prod_{j=1}^{n} K_j \right)^{-1}}{1 + \sum_{i=1}^{n} {l}^{i} \left( \prod_{j=1}^{i} K_j \right)^{-1} }.
\end{eqnarray}
 For example, the exact fraction of product produced by a process of sequential ligands-receptor binding, for a case in which $K_j=K$ and $n=4$ is given by
\begin{align}
    H_d^{S}(l)= \frac{{l}^4}{K^4 + K^3 {l} + K^2 {l}^2 + K {l}^3 + {l}^4}. \label{13}
\end{align}

\begin{figure} [h!t]
\centering
\includegraphics[width=.45\textwidth]{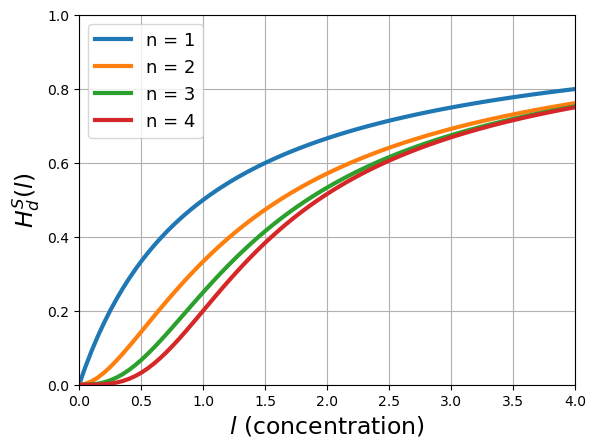}
  \caption{\textbf{Hill function with sequential ligands-receptor binding.} 
This figure shows the behavior of the Hill function where intermediate binding processes occur sequentially, with $K_j=1$ and different values of $n$.}
  \label{fig.2}
\end{figure}

Figure \ref{fig.2} shows the behavior of function (\ref{5.9}) for different values of $n$. 

\subsection{Independent case}
In the case of independent binding, the following reactions and their reaction constants are considered \cite{Weiss}
\begin{equation}
    R + n L \stackbin[k_{-1}]{n *k_{+1}}{\rightleftarrows} RL_{1} + (n-1) L \stackbin[2*k_{-2}]{(n-1)k_{+2}}{\rightleftarrows} RL_{2} + (n-2) L ... \stackbin[n*k_{-n}]{k_{+n}}{\rightleftarrows}RL_{n} , \nonumber
\end{equation}
a similar analysis to the sequential case was performed but with the following reaction rates
\begin{eqnarray}
    R_i^{+}=& k^{+}_i (n-i){l}^{n+1-i} s_{i-1}, \nonumber \\
    R_i^{-}=& k^{-}_i (i) {l}^{n-i} s_{i} ,
\end{eqnarray}
($i=\{1,...,n\}$), where $s_0$ is the concentration of $R$, $l$ is the concentration of $L$ and $s_n$ are the concentrations of $RL_n$. The fraction of product in the reaction as a function of the ligand concentration is given by the next Hill function
\begin{equation}
    H_d^{I}(l)=\frac{{l}^n \left( \prod_{j=1}^{n} K_j \right)^{-1}}{ 1 + \sum_{i=1}^{n} {l}^{i} \frac{n!}{i!(n-i)!} \left( \prod_{j=1}^{i} K_j \right)^{-1}}. 
\end{equation}
An interesting particular case is when $K_j=K$,
\begin{equation}
    H_d^{I}(l)=\frac{l^n}{K^n + \sum_{i=1}^{n} {l}^{i} \frac{n!}{i!(n-i)!} K^{n-i} }= \frac{{l}^n}{(K+{l})^{n}}. \label{16} 
\end{equation}

\begin{figure} [h!t]
\centering
\includegraphics[width=.45\textwidth]{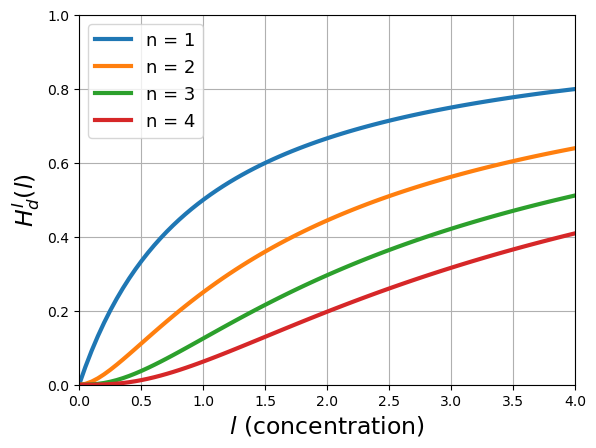}
  \caption{\textbf{Hill function with independent ligands-receptor binding.} This figure shows the behavior of the Hill function where intermediate binding processes occur independently, with $K_j=1$ and different values of $n$.}
  \label{fig.3}
\end{figure}

Figure \ref{fig.3} shows the behavior of the function (\ref{16}) for different values of $n$. As expected, when $n=1$, the Hill function is recovered for both the sequential and independent cases. For the usual Hill function, the Hill coefficient can be better understood as a coefficient that describes the degree of interaction between ligands and receptors, as shown in \cite{Weiss}. 

\section{Intrinsic Fluctuations in Ligands-receptor Reactions}
In addition to the fact that ligands are not solely attached to receptors through a single process, which leads to an effective decimal Hill coefficient, inherent fluctuations can also affect the Hill coefficient value, causing it to deviate from integer values when we fit the experimental data.

\subsection{Sequential case}
In a ligand-receptor process in which sequential binding is required before the final product is produced, the following reactions apply,
\begin{equation}
    R + n L \stackbin[k_{-1}]{k_{+1}}{\rightleftarrows} RL_{1} + (n-1) L \stackbin[k_{-2}]{k_{+2}}{\rightleftarrows} RL_{2} + (n-2) L ... \stackbin[k_{-n}]{k_{+n}}{\rightleftarrows} RL_{n} . \nonumber
\end{equation}
To account for inherent fluctuations, we utilize the expression (\ref{9}), resulting in the following equations for $R_i^+$ and $R_i^-$,
\begin{eqnarray}
    R_i^{+}&= k^{+}_i \left( l^{n+1-i} + \frac{\sigma^2_{l,l}}{2} (n+1-i) (n-i)l^{n-i-1}   \right) s_{i-1}, \nonumber \\
    R_i^{-}&= k^{-}_i \left(l^{n-i} + \frac{\sigma^2_{l,l}}{2}(n-i) (n-i-1)l^{n-i-2}  \right) s_{i},
\end{eqnarray}
where $i=\{1,...,n\}$, $s_0$ represents the concentration of $R$, $l$ denotes the concentration of $L$, and $s_i$ represents the concentration of $RL_i$, stochastic corrections are included in these expressions by the variance between numbers of ligands by $\sigma^2_{l,l}$. In the derivation of these expressions, we assume that the fluctuations between $s_i$ and $l$ are independent; that is, $\sigma^2_{s_i,l}=0$.

In accordance with the stationary condition $R_{i}^{-}=R_{i}^{+}$, we get
{\small
 \begin{eqnarray}
    s_i = s_{i-1} \frac{1}{K_i} \left( \frac{l + \frac{\sigma^2_{l,l}}{2} (n+1-i) (n-i)l^{-1} }{ 1 + \frac{\sigma^2_{l,l}}{2} (n-i) (n-i-1)l^{-2} } \right), \textrm{ with } K_i= \frac{k^{-}_i}{k^{+}_i}. 
\end{eqnarray}}
 Solving for $s_{i}$ in terms of $s_{0}$, we find that
\begin{eqnarray}
    s_i = s_{0} \left( \prod_{j=1}^{i} {K_j}\right)^{-1} l^{i} \left( \frac{l^2  + \frac{\sigma^2_{l,l}}{2}n (n-1) }{ l^2 + \frac{\sigma^2_{l,l}}{2} (n-i) (n-1-i) } \right),
\end{eqnarray}
and finally, the fraction of product in the reaction is given by
\begin{eqnarray}
    H_{sc}^{S}(l)=& \frac{s_n}{s_0 + \sum_{i=1}^{n} s_i} \nonumber \\
    =&\frac{ l^{n-2} \left( {l^2  + \frac{\sigma^2_{l,l}}{2} n (n-1) } \right) \left( \prod_{j=1}^{n} K_j \right)^{-1}}{1 + \sum_{i=1}^{n} l^{i} \left( \frac{l^2  + \frac{\sigma^2_{l,l}}{2} n (n-1) }{ l^2 + \frac{\sigma^2_{l,l}}{2} (n-j) (n-1-i) } \right) \left( \prod_{j=1}^{i} K_j \right)^{-1}}. \label{21}
\end{eqnarray}
The last expression is the Hill-type function with stochastic corrections as it considers the fluctuations in the concentrations of the ligands. For the particular case of $n=4$, $K_j=K$, and $\sigma^2_{l,l} = \frac{l}{\Omega}$ \cite{Manuel}, we obtain
\begin{eqnarray}
       H_{sc}^{S}(l)= \frac{ l^{3} \left( {l  + \frac{ 6}{\Omega} } \right)}{K^4 + K^3 l \left( \frac{l  + \frac{ 6}{\Omega} }{ l + \frac{ 3}{\Omega}  } \right)  + K^2 l^2 \left( \frac{l  + \frac{ 6}{\Omega} }{ l + \frac{ 1}{\Omega}  } \right) + K l^3 \left( \frac{l  + \frac{ 6}{\Omega} }{l} \right) + l^{4} \left( \frac{l  + \frac{ 6}{\Omega} }{l} \right) }. 
\end{eqnarray}
By comparing the last expression with its deterministic counterpart (\ref{13}), it is evident that it differs significantly. Additionally, when $\Omega$ becomes very large, its deterministic counterpart is recovered.

\begin{figure*} [h!t]
\centering
\includegraphics[width=.45\textwidth]{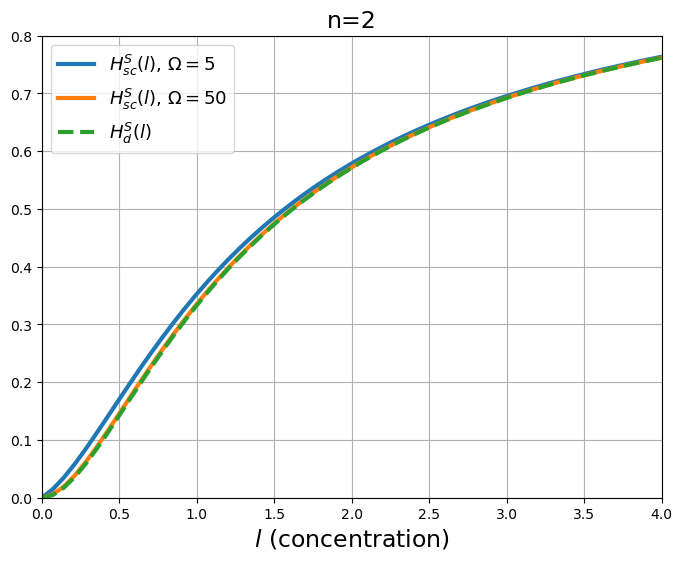}
\includegraphics[width=.45\textwidth]{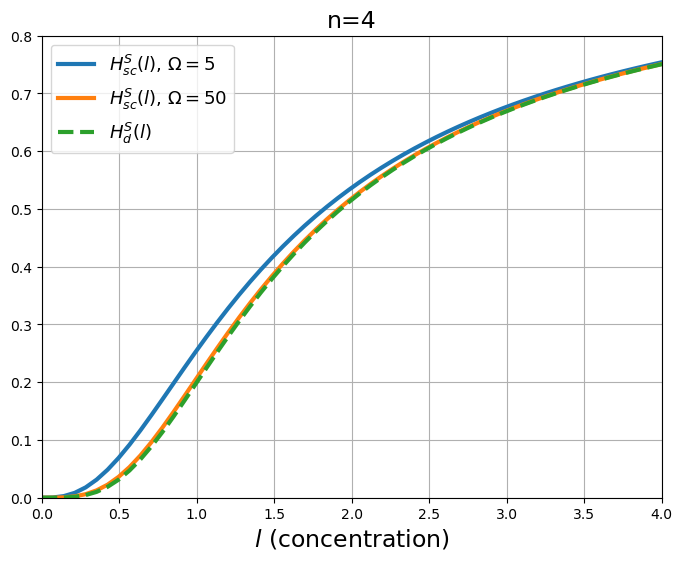}
  \caption{\textbf{Hill function with sequential ligand-receptor binding.} In this figure, we compare a Hill function with sequential intermediate processes in the cases with and without intrinsic fluctuations. We can observe that both functions are similar when $\Omega=50$. We used $K_j=1$  and two different values of $n$. $H_d^{S}(l)$ is the deterministic Hill function and $H_{sc}^{S}(l)$ is the Hill function with stochastic corrections.  }
  \label{fig.4}
\end{figure*}

We present a plot of Equation (\ref{21}) for various values of $n$ and $K_j=K$. The results are shown in Figure. \ref{fig.4}. We compare the deterministic Hill function with intermediate processes $H_d^{S}(l)$ to one that includes stochastic corrections $H_{sc}^{S}(l)$. We plot only two cases, for $n=2$ and $n=4$. The graph indicates that the two functions are very similar when $\Omega=50$ in both cases. However, when $\Omega=5$, there were slight differences between the two, particularly for low concentrations of $l$. Despite these differences, the graphs indicate that the two functions are practically equal when $\Omega$ is big enough.

\subsection{Independent case}
For the independent case, the Hill function with stochastic corrections can be obtained similar to the previous case, yielding the following expression
{\small
\begin{equation}
    H_{sc}^{I}(l)=\frac{l^{n-2} \left( {l^2  + \frac{\sigma^2_{l,l}}{2} n (n-1) } \right) \left( \prod_{j=1}^{n} K_j \right)^{-1} }{ 1 + \sum_{i=1}^{n}  \frac{n!}{i!(n-i)!} \left( \prod_{j=1}^{i} K_j \right)^{-1} l^{i} \left( \frac{l^2  + \frac{\sigma^2_{l,l}}{2} n (n-1) }{ l^2 + \frac{\sigma^2_{l,l}}{2} (n-i) (n-1-i) } \right)}. \label{22}
\end{equation}}
The last expression is a Hill-type function with stochastic corrections because it considers the fluctuations in the ligands. For the case in which $n=4$, $K_j=K$ and $\sigma^2_{l,l}= \frac{l}{\Omega}$, this expression reduces to
\begin{eqnarray}
    H_{sc}^{I}(l)= \frac{l^{3} \left( {l  + \frac{ 6}{\Omega} } \right)}{K^4 + 4 K^3 l \left( \frac{l  + \frac{ 6}{\Omega} }{ l + \frac{ 3}{\Omega}  } \right) + 6 K^2 l^2 \left( \frac{l  + \frac{ 6}{\Omega} }{ l+ \frac{ 1}{\Omega}  } \right) + 4 K l^3  \left( \frac{l  + \frac{ 6}{\Omega} }{ l} \right) + l^{4} \left( \frac{l  + \frac{ 6}{\Omega} }{l} \right) }. \label{23}
\end{eqnarray}
 When $\Omega$ becomes large, its deterministic counterpart is recovered.

\begin{figure*} [h!t]
\centering
\includegraphics[width=.45\textwidth]{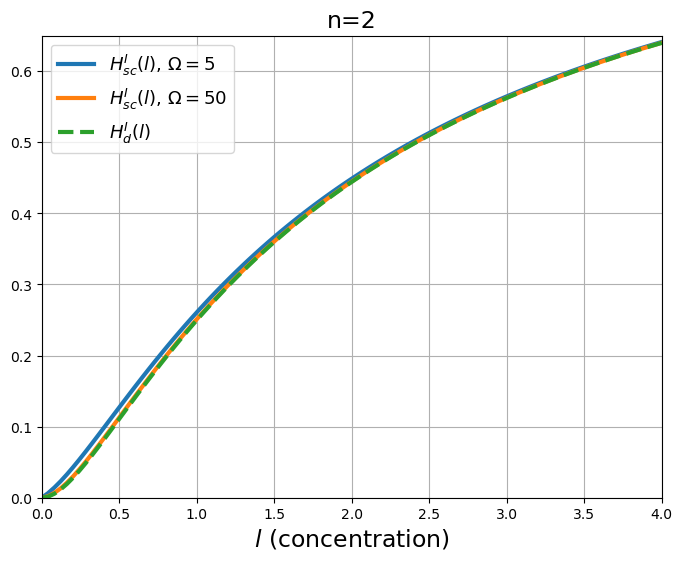}
\includegraphics[width=.45\textwidth]{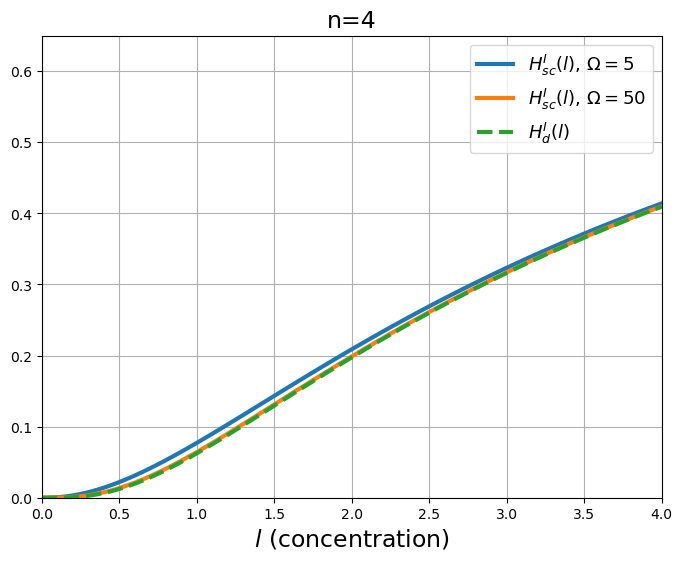}
  \caption{\textbf{Hill function with ligand-receptor binding independent.}In this figure, we compare a Hill function with independently occurring intermediate processes in the cases with and without intrinsic fluctuations. We can observe that both functions almost coincide. We used $K_j=1$, $\Omega=5$ and different values of $n$. With $H_d^{I}(l)$ the deterministic Hill function, and $H_{sc}^{I}(l)$ the Hill function with stochastic corrections.}
  \label{fig.5}
\end{figure*}

In figure \ref{fig.5} we compare the deterministic Hill function with intermediate processes $H_d^{I}(l)$ and the Hill function with stochastic corrections $H_{sc}^{I}(l)$. The figure reveals no discernible differences between the two Hill functions when $\Omega=50$, which suggests that they are practically equivalent. However, when $\Omega=5$ there were few differences when the concentration of $l$ was low.  

For the two cases examined, stochastic corrections are introduced into the system as follows
\begin{eqnarray}
    l^j \rightarrow l^{j} \left( \frac{l^2  + \frac{\sigma^2_{l,l}}{2} n (n-1) }{ l^2 + \frac{\sigma^2_{l,l}}{2} (n-j) (n-1-j) } \right),  \label{rule}
\end{eqnarray}
this transformation represents a term that multiplies the ligand concentration and differs significantly when intermediate processes are not considered. This result differs from what is get in \cite{Manuel}  because here the stochastic corrections are introduced by the following expression,
\begin{equation}
    l^n \rightarrow l^{n} + \frac{\sigma^2_{l,l}}{2} n (n-1) l^{n-2}.
\end{equation}
This shows us that when intermediate processes are considered in the Hill functions, the way in which the corrections are made with respect to the deterministic system differs, and Equation (\ref{rule}) must be used.

The derived Hill-type functions, which incorporate intermediate processes and stochastic corrections, and can be applied to gene regulation networks or to describe enzyme binding. Employing this type of Hill function offers more accurate results as it mitigates the risk of overestimation and enables more precise quantification of the inherent fluctuations within the system. When used in conjunction with the method described in \cite{Manuel} or the Fluctuation Dissipation Theorem (FDT) \cite{Gar}, it enhances the ability to comprehend and model natural variations in the system. 

\section{Relationship Between Intermediate Process and Hill Coefficient.}
This section presents a method for relating empirical Hill functions with decimal coefficients to sequential and independent reaction cases, with and without intrinsic fluctuations.  A specific example is provided to illustrate how they can be related.

Consider a system in which a maximum of $n_m=2$ ligands can bind to a receptor. However, when fitting the experimental data, we obtain $n_e$ (empirical fractional Hill coefficient)  with $1<n_e<2$. By breaking down the Hill function (\ref{H1}), we obtain the following equation
\begin{eqnarray}
    H_d^{f}(l)=& \frac{l^{n_e}}{K^{n_e } + l^{n_e} }, \nonumber \\
   =& \frac{l^{n_m - (n_m-n_e)}}{K^{n_m - (n_m-n_e)} + l^{n_m - (n_m-n_e)}} \nonumber \\
    =& \frac{l^{n_m}}{K^{n_m - (n_m-n_e)}l^{(n_m-n_e)} + l^{n_m} }.  \label{28}
\end{eqnarray}
Next, we can expand $l^{(n_m-n_e)}$ in power series around $l = K$. This choice was made because $H_d^f(l)$, $H_d^S(l)$ (deterministic Hill functions for sequential intermediate processes), and $H_d^I(l)$ (deterministic Hill functions for independent intermediate processes) have the same value at this point \cite{Zsuga}.  Thus, we have
{\small
\begin{equation}
    l^{(n_m-n_e)}= K^{(n_m-n_e)} + (n_m-n_e) K^{(n_m-n_e) - 1} (l-K) + ... ,
\end{equation}}
if we substitute this series in $H_d^f(l)$ then it has a shape similar to the Hill function with an intermediate process.
For example, in the case $n_m=2$ the Hill function can be expressed as follows
\begin{eqnarray}
    H_d^f(l)=& \frac{l^{2}}{K^{2 }(1 + (2-n_e) K^{ - 1} (l-K) + ... ) + l^{2} }  \approx  \frac{l^{2}}{K^{2 } - (2-n_e) K^2 +  (2-n_e) K l  + l^{2} } . \label{31}
\end{eqnarray}
From this result, we can compare the values with the case when considering the intermediate processes. For $n_m=2$, the deterministic Hill functions for sequential and independent intermediate processes are 
\begin{eqnarray}
  H_d^S(l) &= \frac{l^2}{K_1K_2 + K_2 l + l^2}, \nonumber \\
  H_d^I(l) &= \frac{l^2}{K_1K_2 + 2K_2 l + l^2}, \label{32}
\end{eqnarray}
The first equation corresponds to sequential intermediate processes, and the second corresponds to independent intermediate processes. We observe a correspondence between the denominator values in (\ref{31}) and (\ref{32}). This correspondence is presented in the following table.

\begin{table}[h!]
\centering
\begin{tabular}{ |p{2.5cm}|p{3.5cm}|p{3cm}|  }
\hline
\multicolumn{3}{|c|}{Relations for $n_m=2$ and $1<n_e<2$} \\
\hline
Type & $K_1$ & $K_2$ \\
\hline
Sequential & $K(n_e-1)/(2-n_e)$ &  $K(2-n_e)$ \\
Independent & $2K(n_e-1)/(2-n_e)$ &  $K(2-n_e)/2$ \\
\hline
\end{tabular} 
\caption{ \textbf{Relations table without fluctuations.} Relation between the parameter of (\ref{31}) and (\ref{32}) with the deterministic models.}  \label{tabla1}
\end{table}

\begin{figure} [h!t]
  \centering
\includegraphics[width=.45\textwidth]{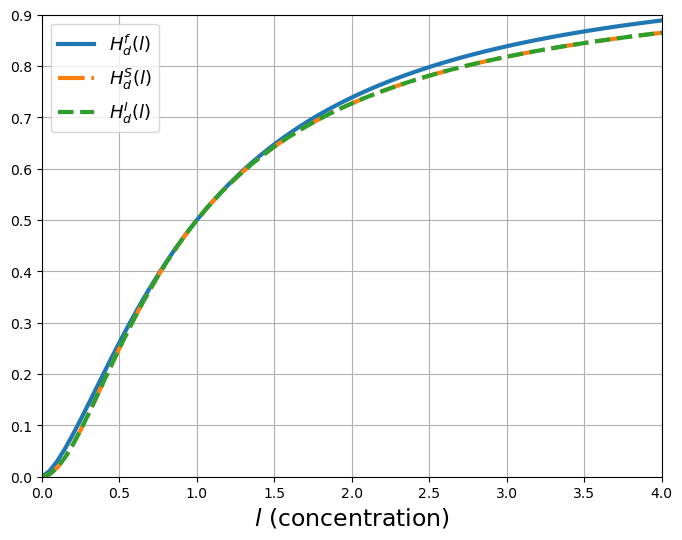}
  \caption{\textbf{Comparison between deterministic models} In this figure, we compare three deterministic models: one using a Hill function with a decimal coefficient ($H_d^f(l)$), another with intermediate processes presented sequentially ($H_d^S(l)$), and the last one with processes presented independently ($H_d^I(l)$). We observe that around $K$ ($K=1$), all three functions overlap. We used $K=1$, $n_m=2$, $n_e=1.5$  and the value of the parameters in the functions are in table \ref{tabla1}.}
  \label{fig.6}
\end{figure}

Figure \ref{fig.6} shows sequential and independent cases as the concentration increases. When the concentrations are close to the value of $K$ ($K=1$), the graphs appear similar but diverge as the concentration increases. Despite the overlapping of the graphs, which makes it difficult to distinguish between sequential and independent cases in this system, it is important to note that they are governed by fundamentally different mechanisms.

From the examples provided, it can be concluded that, when using a Hill function with decimal coefficients, there might be intermediate processes in the reaction. 

Now, we analyze the consequences of intrinsic fluctuations on the Hill coefficient. The Hill functions that involve intermediate processes with intrinsic fluctuations are taken from Equation (\ref{21}) and Equation (\ref{22}), respectively. In the case $n_m=2$, we get 
\begin{eqnarray}
  H_{sc}^{S}(l)=& \frac{{l^2  + \sigma^2_{l,l}  } }{K_1K_2 + K_2  \left( l  + \sigma^2_{l,l}l^{-1} \right) +  \left(  {l^2  + \sigma^2_{l,l}  } \right) }, \nonumber \\
  H_{sc}^{I}(l)=& \frac{ {l^2  + \sigma^2_{l,l}  } }{K_1K_2 + 2K_2 \left( l  + \sigma^2_{l,l}l^{-1}   \right) + \left(  {l^2  + \sigma^2_{l,l}  } \right)}, \label{33}
\end{eqnarray}
where the first equation pertains to sequential intermediate processes and the second pertains to independent intermediate processes. 

We use the rule derived in Equation (\ref{rule}) to introduce fluctuations in a system with intermediate processes, this rule is
\begin{equation}
    l^j \rightarrow l^{j} \left( \frac{l^2  + \frac{\sigma^2_{l,l}}{2} n_m (n_m-1) }{ l^2 + \frac{\sigma^2_{l,l}}{2} (n_m-j) (n_m-1-j) } \right). \label{36}
\end{equation}
Substituting (\ref{36}) into (\ref{31}), results in 
\begin{equation}
    H_{sc}^f(l)= \frac{l^{2} + \sigma^2_{l,l}}{K^{2 } - (2-n_e) K^2 +  (2-n_e) K \left( l  + \sigma^2_{l,l}l^{-1}   \right) + (l^{2} + \sigma^2_{l,l})  } . \label{37}
\end{equation}
It is the Hill-type function with stochastic corrections that can be related to an empirical fractional Hill coefficient $n_{e}$. The system parameters can be recalculated by comparing  (\ref{33}) with (\ref{37}). The results are in Table \ref{tabla4}.

\begin{table}[h!]
\centering
\begin{tabular}{ |p{2.5cm}|p{3.5cm}|p{3cm}|  }
\hline
\multicolumn{3}{|c|}{Relations for $n_m=2$ and $1<n_e<2$} \\
\hline
Type & $K_1$ & $K_2$ \\
\hline
Sequential & $K(n_e-1)/(2-n_e)$ &  $K(2-n_e)$ \\
Independent & $2K(n_e-1)/(2-n_e)$ &  $K(2-n_e)/2$ \\
\hline
\end{tabular} 
\caption{ \textbf{Relations table with fluctuations.}  Relation between the parameter of (\ref{33}) and (\ref{37}) considering intrinsic fluctuations.}  \label{tabla4}
\end{table}

From the outcome of this analysis, it can be inferred that the parameters of the function (\ref{31}), when accounting for stochastic corrections and employing expression (\ref{36}), exhibit an exact alignment with the findings obtained in Table \ref{tabla1}. In other words, the results are consistent with those obtained in the deterministic scenario. Consequently, it is suitable to use (\ref{37}) when considering intermediate processes with intrinsic fluctuations.

\begin{figure} [h!t]
  \centering
\includegraphics[width=.45\textwidth]{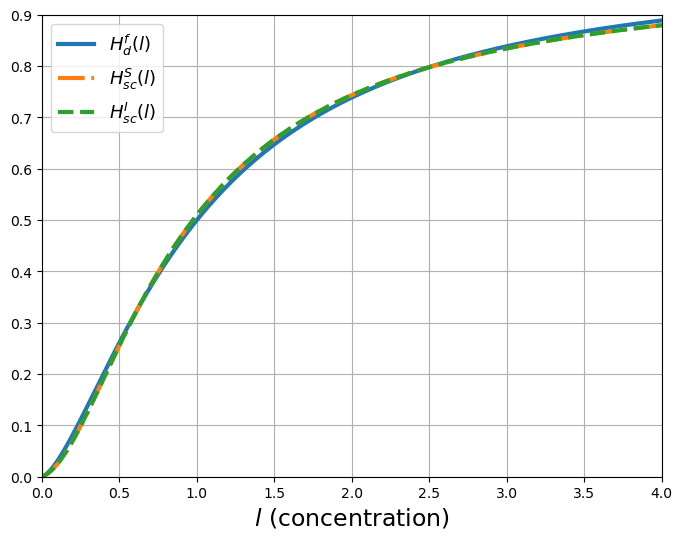}
  \caption{\textbf{Comparison between models with stochastic corrections} In this figure, we adjust the deterministic Hill function with a decimal coefficient ($H_d^f(l)$) using $n_e=1.5$, the Hill function with sequential intermediate processes and stochastic corrections ($H_{sc}^S(l)$), and independent intermediate processes and stochastic corrections ($H_{sc}^I(l)$). We observed that the three functions overlapped and were very similar when the Hill functions with stochastic corrections used $n_e=1.6$ and $\Omega=14.635$. The other parameters used were $n_m=2$, $\sigma^2_{l,l}= \frac{l}{\Omega}$ and $K=1$  and the values of the parameters in the functions are listed in Table \ref{tabla4}.}
  \label{fig.7}
\end{figure}

With this result, the question remains when intrinsic fluctuations in the Hill function with intermediate processes are considered and the value of $n_e$ can change. To do this, we assumed that the data we fitted the deterministic model with a value of $n_e= 1.5$. However, after adjusting the curve generated by the deterministic model using sequential or independent Hill functions with stochastic corrections (we assume $\sigma^2_{l,l}= \frac{l}{\Omega}$ \cite{Manuel}), we found that the values that best reproduced the deterministic Hill function, $H_d^f(l)$, are $\Omega=14.635$ and $n_e=1.6$, as shown in Figure \ref{fig.7} (we set $K=1$). 

Thus, we can conclude that by introducing intrinsic fluctuations into the system, we can better adjust the system using models with stochastic corrections as well as by changing the values of $n_e$. Models with intermediate processes and stochastic corrections are more realistic because the ligands do not bind instantly to the receptors, and fluctuations are always present.

\section{Conclusions}
  A more general expression for the Hill function can be obtained by considering intermediate and stochastic processes.  This generalization of the Hill function can help improve the accuracy of the description of biochemical network dynamics when intermediate processes are present and in the mesoscopic regime. This type of description can be useful for simulating networks with a large number of nodes because it is computationally more efficient than the commonly used Gillespiee algorithm. This opens the door to make simulations of mesoscopic systems with a large number of components that are more closely related to real-world scenarios than other simple models. By establishing a connection between the intermediate processes and the Hill function with decimal coefficients, it was possible to show that the Hill function with a decimal coefficient is equivalent to describing a process with intermediate processes. Furthermore, when intrinsic fluctuations are introduced to the system and a Hill function is parameterized with stochastic corrections, the decimal Hill coefficient varies with respect to that obtained deterministically, showing that fluctuations can play an important role when calculating this coefficient. This relationship provides a better understanding of the underlying processes associated with the decimal Hill coefficient while also enabling the prediction of an effective value of the Hill coefficient from the underlying mechanism, allowing us to have a simplified description of complex systems.

\section*{Acknowledgments}
\noindent The author Manuel E. Hernández-García acknowledge the financial support of CONAHCYT through the program "Becas Nacionales 2023".\\
Jorge Velázquez-Castro acknowledge financial support of VIEP-BUAP through project 00226-VIEP 2023.

\section*{Declarations}
The authors declare no conflicts of interest regarding the publication of this article. 

All data generated or analyzed in this study are included in this published article.

\end{document}